\documentclass[epj]{svjour}
\usepackage{latexsym}
\usepackage{graphicx}

\begin{document}

\title
{Spin-dependent conductance statistics in systems with spin-orbit coupling}

\author{S L A de Queiroz}
\institute{Instituto de F\'\i sica, Universidade Federal do
Rio de Janeiro, Caixa Postal 68528, 21941-972
Rio de Janeiro RJ, Brazil}
\date{Received: date / Revised version: date}
%
\abstract{
Spin-dependent partial conductances are evaluated in 
a tight-binding description of electron transport in the presence 
of spin-orbit (SO) couplings, using transfer-matrix methods.
As the magnitude of SO interactions increases, the separation
of spin-switching channels from non-spin-switching ones is gradually
erased. Spin-polarised incident beams are produced by
including a Zeeman-like term in the Hamiltonian. The exiting
polarisation is shown to exhibit a maximum as a function of the intensity
of SO couplings. For moderate site disorder, and both weak and strong
SO interactions, no evidence is found for a decay of exiting
polarisation against increasing system length. With very low site
disorder and weak SO couplings a spin-filter effect
takes place, as polarisation {\em increases} with increasing
system length.
} 
\maketitle

\section{Introduction}
\label{intro}

In this paper we study the statistics of the direct-current (DC) conductance
 of two-dimensional (2D) systems with spin-orbit (SO) 
interactions~\cite{hln80,b84,ando89}, in particular its spin-dependent 
properties and their dependence on various ex\-ternally-imposed parameters.

Investigation of the interplay between the electron's spin magnetic moment 
and  assorted  properties of electronic systems is the subject of 
the emerging field of spintronics~\cite{zutic,bader,baltz}. 
Here we start from a tight-binding, spin-dependent Hamiltonian.
Although this is a one-electron description, it is of course a full 
quantum-mechanical one. Thus the exclusion effects associated
with Pauli's principle are implicit in the anticommutation rules
obeyed by the respective single-particle creation and annihilation 
operators involved. In this context, our main aim is to give a
quantitative account of the joint effects of exclusion, on the one hand, 
and spin flipping (enabled by SO couplings), on the other,
upon electronic transport in disordered systems. Qualitatively,
one will be asking whether an electron driven across a sample by a 
weak electric  field will preferentially maintain the spin with which 
it was injected, or revert it via SO interactions; also of interest is 
the total current flowing through the sample in such circumstances.

We compare selected results of our calculations, especially those 
pertaining to spin polarisation, to those obtained for a variant of 
the similar (though classical) problem of directed flow with exclusion
which mimicks the Pauli principle by allowing for conditional
site double-occupation~\cite{dqrbs17}.

Section~\ref{sec:th} below recalls some technical aspects of the 
transfer-matrix (TM) method used in our
calculations. In Sec.~\ref{sec:num} we give numerical results for the
statistics of spin-dependent conductances, both for spin-unpolarised
systems and for the polarised case. For the latter we also investigate the
behaviour of the polarisation itself, and of spin-correlation
functions. In Sec.~\ref{sec:conc} we summarize and discuss
our results.

\section{Theory}
\label{sec:th}

The model one-electron Hamiltonian for this problem is
\begin{equation}
{\cal H}= \sum_{i,\sigma} \varepsilon_i\,c^\dagger_{i\sigma}\,c_{i\sigma} +
\sum_{\langle i,j\rangle}\,\sum_{\sigma, \sigma^\prime} V_{ij\,\sigma\sigma^\prime}
\,c^\dagger_{i\sigma}\,c_{j\sigma^\prime}\ +\ {\rm h.c.}\ ,\!\!
\label{eq:hdef}
\end{equation}
where $c^\dagger_{i\sigma}$, $c_{i\sigma}$ are creation and annihilation operators
for a particle with spin eigenvalues $\sigma=\pm1$ at site $i$, and the
self-energies
$\varepsilon_i$ are  independently-distributed random variables;
$V_{ij\,\sigma\sigma^\prime}$ denotes the $2 \times 2$ spin-dependent hopping
matrix between pairs of nearest-neighbour sites $\langle i,j\rangle$,
whose elements must be consistent with the symplectic symmetry of SO
interactions~\cite{mc92,mjh98}.
In Eq.~(\ref{eq:hdef}) we describe SO couplings via an effective Hamiltonian
with a  single ($s-$like) orbital per site~\cite{ando89}.
Several possible forms may be considered for the hopping term, depending on whether
one is specifically considering Rashba- or Dresselhaus- like couplings~\cite{ando89,kka08},
or (as is the case here) the focus is simply  on the basic properties of systems in the
symplectic universality class~\cite{ez87,e95,aso04}.

Here we use the implementation of~\cite{ez87,e95} for the hopping term, namely:
\begin{equation}
V_{ij}=I+\mu i\!\sum_{k=x,y,z}\!V_{ij}^k\sigma^k=\!
\left(\matrix{\!1+i\mu V_{ij}^z&\!\mu V_{ij}^y+i\mu V_{ij}^x\cr
\!-\mu V_{ij}^y+i\mu V_{ij}^x&\!1-i\mu V_{ij}^z\cr}\right)\!,
\label{eq:vdef}
\end{equation}
where $I$ is the $2\times 2$ identity matrix, $\sigma^k$ are the Pauli
matrices, and $\mu$ gives the intensity of the SO coupling;
below we consider the (real) $\{V_{ij}^k\}$
uniformly distributed in $[-1/2,1/2]$. Thus
all energies are written in units of the $\mu \equiv 0$ nearest-neighbour
hopping.

The $\varepsilon_i$ are taken from a random uniform distribution 
in $[-W/2,W/2]$.

The form Eq.~(\ref{eq:vdef}) for the hopping term does not exhibit the
explicit multiplicative coupling between momentum and spin degrees of freedom,
characteristic of Rashba-like Hamiltonians~\cite{ando89,kka08}.
In two dimensions one should not expect significant discrepancies between
results from either type of approach, as long as one is treating systems without
lateral confinement.

Since we shall not attempt detailed numerical comparisons to experimental
data, the simplified formulation described above seems adequate
for our purposes.

We apply the TM approach specific to tight-binding
Hamiltonians like Eq.~(\ref{eq:hdef})~\cite{ps81,ps81b},
with suitable  adaptations 
for incorporating SO effects~\cite{ynik13},
considering a strip of the square lattice cut along one of the 
coordinate directions with periodic boundary conditions across.
With SO couplings along the bonds, the 
site-dependent wavefunction amplitudes $a_{ik}$ ($i$,$k$ being site
coordinates) are now
spinors, written on the basis of the eigenvectors of $\sigma^z$ as:
\begin{equation}
a_{ik} = \left(\matrix{a_{ik}^{\uparrow}\cr a_{ik}^{\downarrow}\cr}\right)\ .
\label{eq:spinor}
\end{equation}

Our calculations of the two-terminal DC conductance follow the procedure 
described in~\cite{pmr92}, in which the TM is iterated and projected in such a way that the
transmissivity matrix $t$ can be retrieved~\cite{pmr92,tm02,nazarov}. From that the zero-field
conductance $g$ is evaluated by the Landauer formula~\cite{nazarov,l70}
\begin{equation}
g = {\mathrm Tr}\ tt^\dagger = \sum_{i,j} |t_{ij}|^2\ ,
\label{eq:cond}
\end{equation}
where the sum runs over all entry ($j$) and exit ($i$) channels.

In this scheme one considers a disordered system described by Eq.~(\ref{eq:hdef}),
with $N$ sites across and length $M$ sites, connected to pure leads at both ends (that is,
where all $\varepsilon_i \equiv 0$, $\mu \equiv 0$). Eq.~(\ref{eq:cond})
gives the {\em total} conductance, i.e., considering both spin directions. 
The methods of~\cite{pmr92} were used to evaluate the probability distribution of the total
conductance~\cite{rms01,markos02,osk04,ms06}, at the metal-insulator transition for
2D systems with SO couplings.

For spin-dependent properties one considers the partial conductances
$g_{\sigma\sigma^\prime}$, where
\begin{equation}
g_{\sigma\sigma^\prime}= \sum 
\left| t_{\sigma\sigma^\prime}\right|^2\ ,
\label{eq:gab}
\end{equation}
with  $t_{\sigma\sigma^\prime}$ being the transmission coefficient from
the left lead with spin $\sigma^\prime$ to the right lead with spin
$\sigma$~\cite{ohe03,ohe05}. The sum runs over all such pairs of channels
with fixed $\sigma$, $\sigma^\prime$.

\section{Numerics}
\label{sec:num}

\subsection{Introduction}
\label{sec:num-intro}

Here we take $E=0$, corresponding to the Fermi energy
of a non-disordered system with  $W=\mu \equiv 0$.
At $E=0$, $\mu=2.0$ the critical disorder for the metal--insulator
transition on a square lattice is estimated as $W_c=8.55\,(5)$~\cite{e95}.

For each set of $E$, $\mu$, and $W$ studied
we generally take $N_s= 10^5$ independent realizations of
disorder, which enables us to obtain smooth curves for the
distributions of conductances and associated quantities.

\subsection{Unpolarised beams}
\label{sec:unpol}

The plane wave-like incident states used are straightforward adaptations
of those used for the spinless case~\cite{pmr92}, with the spinors of
Eq.~(\ref{eq:spinor}) having equal components. So in this case the incident
beam is fully spin-unpolarised.

\begin{figure}
\begin{center}
\hskip-.7cm
\includegraphics[width=3.6in]{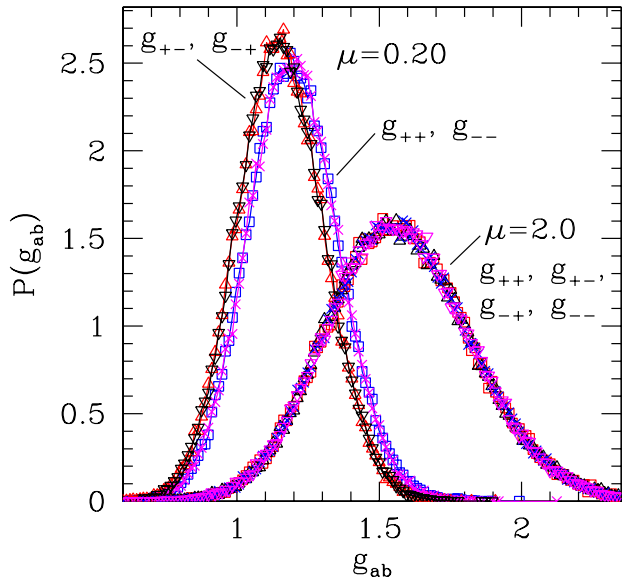}
\end{center}
\vskip-1.2cm
\caption{
Probability distributions for  partial conductances
$g_{ab}$, $a,b=\pm$ of system with $M=N=40$ sites for energy
$E=0$, $W=3.0$, with $\mu=0.20$ (leftmost curves) and $2.0$,
see Eqs.~(\ref{eq:hdef}) and~(\ref{eq:vdef}), both in the metallic phase.
The incident beams are spin-unpolarised.
$N_s=10^5$ independent samples for each case.
}
\label{fig:p0_gab_c}
\end{figure}

We now specialise to the conducting phase.
With $E=0$, $\mu=2.0$ we took $W=3$, well within the metallic
phase. Then, keeping $W$ fixed we lowered the SO coupling by an order
of magnitude, making $\mu=0.2$. For this $\mu$ the critical disorder
at the centre of the unperturbed band is $W_c \approx 5.3$.
The results for the distributions of the $g_{ab}$ are displayed in 
Fig.~\ref{fig:p0_gab_c}, confirming to a very good extent their
expected Gaussian shape~\cite{markos02,jlp91}.

While for $\mu=2.0$ all four curves
coincide (we checked that this also happens at the metal-insulator transition
for this $\mu$, namely $W_c=8.55$), there is
a split between the spin-conserving and spin-flipping 
groups at $\mu=0.20$ where the SO effects are weaker. For the lower $\mu$
the peaks of the $(+-),(-+)$ group
are located at a value of $g$ some $3.7\%$ less than those of $(++),(--)$.
For comparison, for $\mu=2.0$ the (non-systematic) spread among the peak
locations for all four distributions is of order $0.3\%$.

\subsection{Spin-polarised beams}
\label{sec:spol}

When attempting to represent spin-polarised
mixtures of spins with the methods of~\cite{pmr92},
the following procedures may be devised in order to set a specific incoming
polarisation: either (i) impose unequal amplitudes for $\uparrow$
and $\downarrow$ spinor components on
the input state vectors, see Eq.~(\ref{eq:spinor}),
or (ii) suitably modify the Hamiltonian Eq.~(\ref{eq:hdef}) while keeping the
input vectors unconstrained.

We have considered procedure (i); this corresponds to a straightforward
generalization of well-known procedures used in variants of
the classical problem of directed flow with exclusion,
which include conditional allowance of site
double-occupation (mimicking the Pauli principle)~\cite{dqrbs17}.
Overall, we found that the effect of using
such constrained input vectors amounts to a modulation of
the channel-specific conductances by the respective incoming and
outgoing polarisations, the latter being predicted to equal the
former on the basis of the time-reversal symmetry of
the Hamiltonian~\cite{nazarov}. Though numerical data have been
found to be compatible with such conservation, 
it is not clear at this point how one might extract additional
physical insight from the resulting picture. 

We thus concentrated on procedure (ii).
One can add spin-dependent energies to Hamiltonian Eq.~(\ref{eq:hdef}),
{\em e.g.}, by introducing an external
magnetic field as done in~\cite{ohe03} in a similar context.
In this case, at energy $E$ there are more open propagation
channels ($\varepsilon_i (\sigma) > E$) for one spin species than
for the other.
Analysis of level-spacing statistics has shown that,
upon application of such field, the universality class of the problem
changes from the Gaussian symplectic ensemble  to that
of the Gaussian unitary ensemble~\cite{ohe03}. This is to be expected
from symmetry considerations~\cite{been97}.
Furthermore, conservation of average polarisation
is now not guaranteed since time-reversal symmetry has been 
broken by the magnetic field~\cite{nazarov}.

 At least two possible implementations can be suggested: (i) a field 
acting only on the left lead~\cite{ohe03},
or (ii) a spin-dependent chemical potential acting on every site
of the sample. For simplicity here we stick to  option (i).

With $\varepsilon_i^0$ being the self-energy of an orbital on the
left lead in the absence of field, we establish our
Zeeman-like energy origin such that $\varepsilon_i (+)=\varepsilon_i^0$,
$\varepsilon_i (-)=\varepsilon_i^0+\Delta$ ($\Delta >0$),
so propagation of plus spins is favored. Below we make $\Delta=3.75$.
As the unperturbed single-electron band spans the interval
$-4 \leq E \leq 4$ in the square-lattice geometry, this induces a 
significant spin imbalance. We also considered other values in the
range $1.0 \leq \Delta \leq 4.0$, with qualitatively similar
results.

Here we always take the incoming beams as fully
spin-unpolarised, so any resulting polarisation is driven by the
Zeeman-like gap $\Delta$.

As in Section~\ref{sec:unpol} we specialize to the conducting phase, 
so the conductance distributions are expected to be close to
Gaussian~\cite{markos02,jlp91}. In this context the central estimates,
denoted by $\langle \cdots \rangle$,
and error bars given for conductances and related quantities below
correspond, respectively, to mean value and RMS deviation
of the associated distributions.

\begin{figure}
\begin{center}
\hskip-.7cm
\includegraphics[width=3.6in]{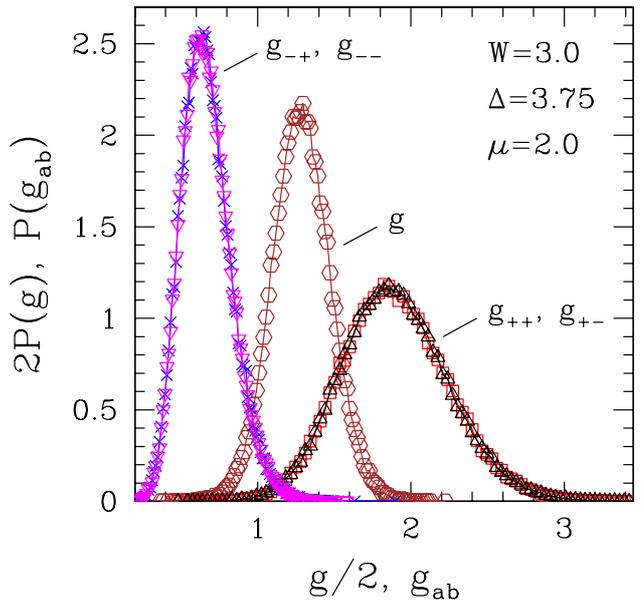}
\end{center}
\vskip-1.2cm
\caption{
Probability distributions for total ($g$) and partial ($g_{ab}$, with
$a,b=\pm$) conductances of system with $M=N=20$ sites for energy
$E=0$, in the conducting phase with $\mu=2.0$, $W=3.0$, and $\Delta=3.75$.
$N_s=10^5$ independent samples.
}
\label{fig:del375_mu2_m20}
\end{figure}

\begin{figure}
\begin{center}
\hskip-.7cm
\includegraphics[width=3.6in]{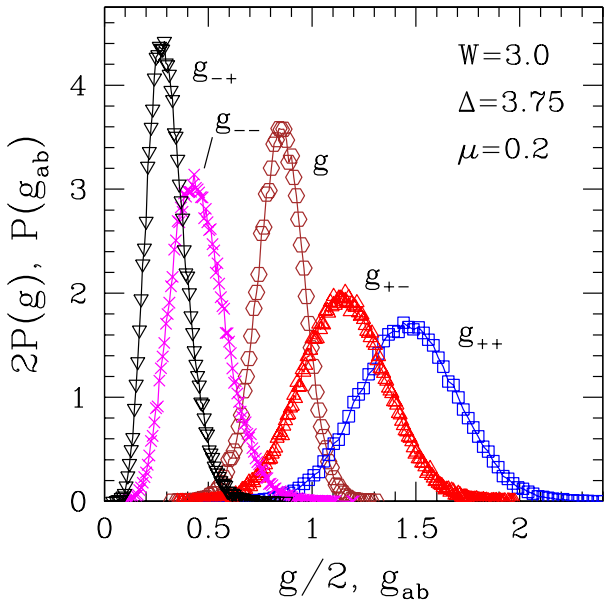}
\end{center}
\vskip-1.2cm
\caption{
Probability distributions for total ($g$) and partial ($g_{ab}$, with
$a,b= \pm$) conductances of system with $M=N=20$ sites for energy
$E=0$, in the conducting phase with $\mu=0.2$, $W=3.0$, and $\Delta=3.75$.
$N_s=10^5$ independent samples.
}
\label{fig:del375_mu02_m20}
\end{figure}

Figs.~\ref{fig:del375_mu2_m20}
and~\ref{fig:del375_mu02_m20} show the distributions for total
and partial conductances, and should be compared, respectively,
with the rightmost and leftmost groups of
Fig.~\ref{fig:p0_gab_c}, which both correspond to $\Delta= 0$.
For $\mu=2.0$ (Fig.~\ref{fig:del375_mu2_m20}) channel asymmetry
is manifest in that there is a grouping of distributions by exiting
channels, respectively for majority $[\,(+,+)$ and $(+,-)\,]$ and
minority  $[\,(-,+)$ and $(-,-)\,]$ spins, while for $\Delta=0$ all four
distributions coincide. Note that this is {\em not} the same
pairing exhibited by the $\mu=0.2$ curves in Fig.~\ref{fig:p0_gab_c},
where the groupings are spin-flipping $[\,(+,-)$ and $(-,+)\,]$ and 
spin-conserving $[\,(+,+)$ and $(-,-)\,]$, both due to up-down symmetry.

For $\mu=0.2$ (Fig.~\ref{fig:del375_mu02_m20}) all four channels
are fully split, as up-down symmetry is now absent. 
However, a tendency towards the same grouping of 
Fig.~\ref{fig:del375_mu2_m20} still holds.

\begin{figure}
\begin{center}
\hskip-.7cm
\includegraphics[width=3.6in]{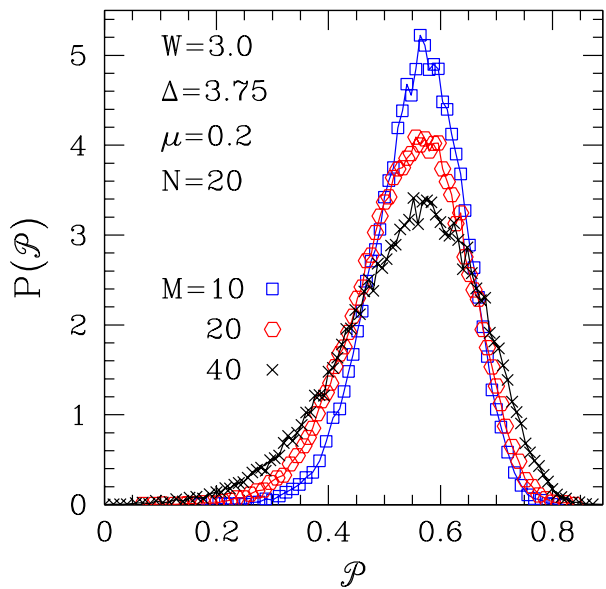}
\end{center}
\vskip-1.2cm
\caption{
Probability distributions for exiting polarisation ${\cal P}$ at
right end of systems of fixed width $N=20$ sites, and varying length
$M=10$, $20$, and $40$ sites. All for $E=0$, in the conducting phase
with $\mu=0.2$, $W=3.0$, and $\Delta=3.75$.
$N_s=10^5$ independent samples.
}
\label{fig:pol_mu02_mvar}
\end{figure}
The polarisation $\cal P$ of the exiting current can be
evaluated~\cite{ohe05} as
\begin{equation}
{\cal P}= \frac{g_{++}+g_{+-}-(g_{--}+g_{-+})}{g_{++}+g_{+-}+g_{--}+g_{-+}}\ ,
\label{eq:pol_def}
\end{equation}
see Eq.~(\ref{eq:gab}).

In Fig.~\ref{fig:pol_mu02_mvar} the probability distributions for $\cal P$
are shown, for systems with fixed width and varying aspect ratios $M/N$.
The distributions broaden out as the system becomes
more wire-like, and develop a more pronounced negative skew. On the
other hand the position of their peak value remains essentially unchanged.
So, despite the Hamiltonian not being invariant under time reversal,
we see indications that polarisation of the beam is conserved in this 
case. 

While Fig.~\ref{fig:pol_mu02_mvar} is for $\mu=0.2$, we have found
qualitatively very similar outcomes for larger $\mu$, in particular the
stability of the peak location for $P({\cal P})$ against varying $M$.
Having this in mind, we kept $M=10$ fixed and varied several
other parameters. In this way we are reasonably sure to
retain the main qualitative features of the quantities under
study, while the spread of distributions is kept lower than for larger $M$.

\begin{figure}
\begin{center}
\hskip-.7cm
\includegraphics[width=3.6in]{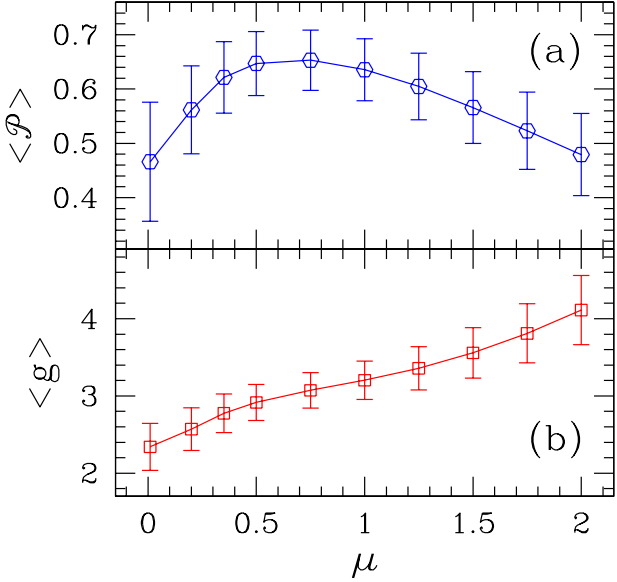}
\end{center}
\vskip-1.2cm
\caption{
For $E=0$ with $W=3.0$, $\Delta=3.75$, and varying $\mu$,
panel (a) shows average exiting polarisation $\langle{\cal P}\rangle$
at right end, and (b) gives average total conductance $\langle g \rangle$,
of systems of fixed width $N=20$, length $M=10$ sites.
All for $N_s=10^5$ independent samples.
}
\label{fig:pol_cond_375}
\end{figure}

Fig.~\ref{fig:pol_cond_375} shows, for fixed $\Delta=3.75$, the variation
of the average exiting polarisation $\langle{\cal P}\rangle$, and total
conductance $\langle g \rangle$, against
the  intensity $\mu$ of SO coupling. The maximum exhibited in panel (a)
for $\mu \approx 0.75$ will be discussed in detail below.
Note however that the difference between
minimum and maximum values of $\langle {\cal P} \rangle$ is only some
$30\%$ of the largest, on account of the largish value of $\Delta$.
On the other hand, panel (b) shows that
increasing SO effects lead to monotonically increasing
total conductance, for fixed $\Delta$.

The leftmost points of panel (a), (b) of Fig.~\ref{fig:pol_cond_375}
correspond to $\mu=0.01$ which most probably is in the insulating
phase; approximate calculations indicate that for $W=3.0$ the
metal-insulator boundary lies at $\mu \approx 0.06$. Even so,
such point in $(W, \mu)$ space would have an associated localization
length much larger than the current system's size. Thus we believe it
is justifiable to present data for this point together with
those unequivocally belonging in the conducting phase.

The unusual behaviour exhibited by $\langle{\cal P}\rangle$ against $\mu$
 warrants further investigation. In order to  check on the possible 
influence of finite-size effects we ran simulations
for $N=40$ and $60$, keeping $M=N/2$ to preserve the aspect ratio,
and spanning the same set of $\mu$ values as in Fig.~\ref{fig:pol_cond_375}.
Except for very weak SO coupling, $\mu=0.01$, where the centre of the 
distribution of $\cal P$ shifts to smaller values by a significant amount
upon increasing $N$, 
we see no  shift for the corresponding centres at larger $\mu$ as $N$ increases 
(in which case the only noticeable effect is that the peaks become sharper).
In conclusion, the non-monotonic 
variation of $\langle{\cal P}\rangle$ against $\mu$ observed in 
Fig.~\ref{fig:pol_cond_375} does not appear to be an artifact from finite 
sample dimensions.

To see the physical origin of the polarisation maximum one must first 
follow the evolution, for fixed exit channel, of the split between 
spin-conserving and spin-flipping partial  conductances; then one
must consider how the {\em difference} between said splits evolves.
We define
\begin{equation}
\delta g_+ \equiv g_{++} - g_{+-} \ ;\quad \delta g_- \equiv g_{--} - g_{-+} \ ,
\label{eq:gsplit}
\end{equation}
where the $\mu$-- dependence of all quantities is implicit.

Firstly, both $\delta g_+$ and $\delta g_-$ are positive
at small $\mu$ (see Fig.~\ref{fig:del375_mu02_m20}), and vanish
upon increasing $\mu$ (see Fig.~\ref{fig:del375_mu2_m20}).
Secondly, for $\Delta >0$ one has
$\delta g_+ > \delta g_-$ (see Fig.~\ref{fig:del375_mu02_m20} for an
example), because: (i) the external field generally favours
upward-pointing spins, so $g_{++} > g_{--}$, while (ii) for any $\mu$ and for 
fixed exit channel, the ratio of spin-flipping to spin-conserving conductance
is nearly independent of exit channel, as will be
 shown in panel (b) of Fig.~\ref{fig:pol_comp} below. 
Eq.~(\ref{eq:pol_def}) can be rewritten as
\begin{equation}
{\cal P}= \frac{2(g_{++}-g_{--})}{\Sigma} +
\frac{(\delta g_--\delta  g_+)}{\Sigma}\ ,
\label{eq:pol_def2}
\end{equation}
with $\Sigma \equiv g_{++}+g_{+-}+g_{--}+g_{-+}$. 
\vskip.1cm\par
\begin{figure}
\begin{center}
\hskip-.7cm
\includegraphics[width=3.6in]{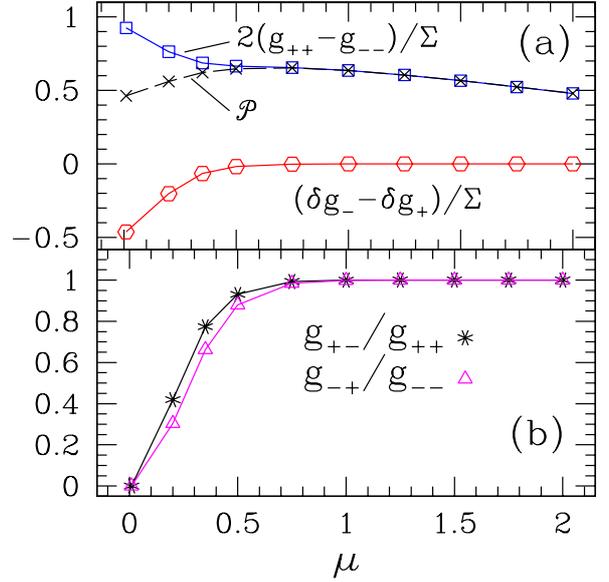}
\end{center}
\vskip-1.2cm
\caption{
Panel (a): plots of the individual contributions from each of the terms on the 
right-hand side of Eq.~(\ref{eq:pol_def2}).
The resulting polarisation $\cal P$ is shown (crosses joined by dashed line).
Panel (b): ratios of spin-flipping to spin-conserving conductances according
to exit channel.
Error bars are omitted for clarity.
All for $E=0$ with $W=3.0$, $\Delta=3.75$, $N=20$, $M=10$, 
$N_s=10^5$ independent samples.
}
\label{fig:pol_comp}
\end{figure}

Panel (a) of Fig.~\ref{fig:pol_comp} shows how each of the two terms 
on the right-hand 
side of  Eq.~(\ref{eq:pol_def2}) behaves individually against $\mu$.
The contribution given by the first (spin-conserving) one follows the 
expected decreasing trend upon increasing $\mu$, in accord  
with the idea that SO 
effects tend to equalise up- and downward-pointing spins.  On the other hand,
both $\delta g_+$ and $\delta g_-$ vanish for $\mu \approx 0.75$ (see also 
panel (b) of Fig.~\ref{fig:pol_comp}).  Clearly the initially-growing 
behaviour of $\cal P$ is due to the negative, and
diminishing, contribution of the second term. 

We now turn to cases in which the effects of site disorder are suppressed.
Still with $E=0$ and $\Delta=3.75$ we took $W=10^{-5}$, which
is essentially zero for practical purposes here.
We considered $P({\cal P})$ for systems of width $N=20$ sites
across, and lengths $M=10$, $20$, and $40$. The picture for $\mu=2.0$
was qualitatively and quantitavely very similar to that given in
Fig.~\ref{fig:pol_mu02_mvar} for $W=3.0$, $\mu=0.2$: peak positions
and RMS widths respectively  $0.54-0.55$ and $0.04-0.06$ ($W=10^{-5}$, $\mu=2.0$),
$0.54-0.56$ and $0.08-0.12$ ($W=3.0$, $\mu=0.2$).

\begin{figure}
\begin{center}
\hskip-.7cm
\includegraphics[width=3.6in]{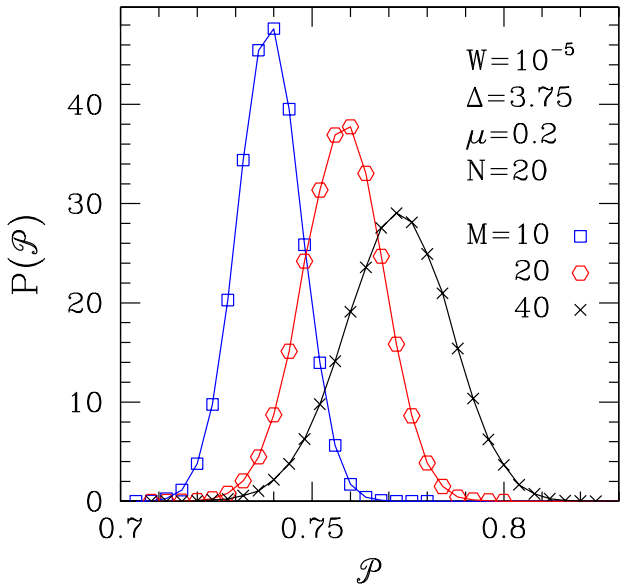}
\end{center}
\vskip-1.2cm
\caption{
Probability distributions for exiting polarisation ${\cal P}$ at
right end of systems of fixed width $N=20$ sites, and varying length
$M=10$, $20$, and $40$ sites. All for $E=0$, in the conducting phase
with $\mu=0.2$, $W=10^{-5}$, and $\Delta=3.75$.
$N_s=10^5$ independent samples.
}
\label{fig:pol_mu02_w0_mvar}
\end{figure}

On the other hand, for weak SO couplings $\mu=0.2$, as shown in
Fig.~\ref{fig:pol_mu02_w0_mvar}, it is seen that for this
combination of parameters the system can function as a
spin {\em filter}~\cite{ohe05}, with the beam polarisation increasing as
it propagates. Quantitatively,
$\langle{\cal P}\rangle(M=40)=0.774(14)$,
$\langle{\cal P}\rangle(M=10)=0.741(8)$. 

Taking strong SO coupling $\mu=2.0$, the spin-filter effect becomes
essentially undiscernible: one gets $\langle{\cal P}\rangle(M=40)=0.541(75)$,
$\langle{\cal P}\rangle(M=20)=0.540(65)$,
$\langle{\cal P}\rangle(M=10)=0.541(55)$.

In this region of parameter space one can capture the decay of the
spin-spin correlation function~\cite{kka08}
\begin{equation}
F_{zz}(L)= \langle \sigma^z(0)\,\sigma^z(L)\rangle\ ,
\label{eq:fzz}
\end{equation}
which is the conditional probability to find an electron with spin $\sigma^z$
at $x=L$, for an incident electron at $x=0$ with $\sigma^z=\pm 1$.
In terms of the $g_{\sigma \sigma^\prime}$ of Eq.~(\ref{eq:gab}), this is:
\begin{equation}
F_{zz} = \frac{1}{2}\,{\Big\langle}
\frac{g_{++}-g_{-+}}{g_{++}+g_{-+}}+\frac{g_{--}-g_{+-}}{g_{--}+g_{+-}}
{\Big\rangle}\ .
\label{eq:fzz2}
\end{equation}

\begin{figure}
\begin{center}
\hskip-.7cm
\includegraphics[width=3.6in]{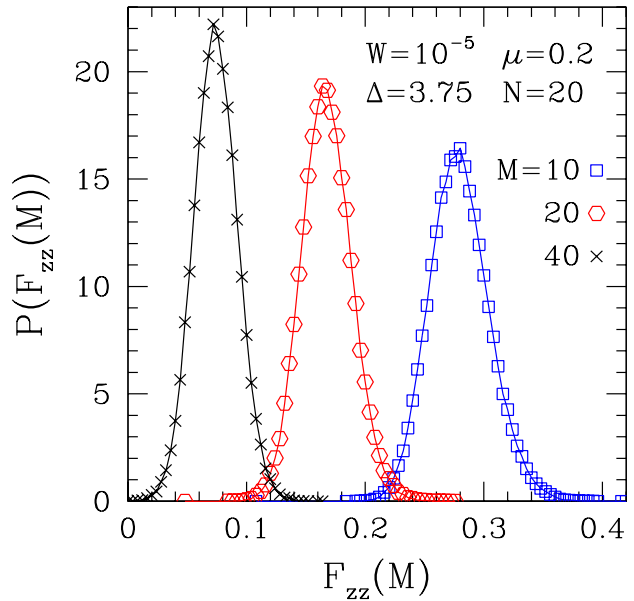}
\end{center}
\vskip-1.2cm
\caption{
Probability distributions for correlation function $F_{zz}(M)$ [$\,$see
Eqs.~(\ref{eq:fzz}),~(\ref{eq:fzz2})$\,$] between left and
right end of systems of fixed width $N=20$ sites, and varying length
$M=10$, $20$, and $40$ sites. All for $E=0$, in the conducting phase
with $\mu=0.2$, $W=10^{-5}$, and $\Delta=3.75$.
$N_s=10^5$ independent samples.
}
\label{fig:corf_mu02_w0_mvar}
\end{figure}

The spin relaxation length $\Lambda_s$~\cite{kka08} is defined by assuming
an exponential decay for the correlation function, $F_{zz}(L) \propto
\exp (-L/\Lambda_s)$. Fig.~\ref{fig:corf_mu02_w0_mvar} displays data
for $E=0$, $\Delta=3.75$, $W=10^{-5}$, $\mu=0.2$, from which a fit to
the above expression gives $\Lambda_s=22(2)$. Of course this is intended
mostly for illustration, since here one has only three data points.
Nevertheless, a semi-quantitative analysis of such results can prove
enlightening. For example, keeping the same set of parameters and system
sizes but making $W=3.0$ gives $\Lambda_s=9.0(2)$; if one then keeps
$W=3.0$ and makes $1.0 \leq \Delta \leq 4.0$, $\Lambda_s$ remains within
$10\%$ of that. However, for strong SO coupling $\mu=2.0$, spin relaxation
takes place at very short distances even for very low site disorder.
In this case, already for $M=10$ the distributions of $F_{zz}(M)$
remain essentially centred around zero for all sets of physically
plausible parameters used here.

\section{Discussion and Conclusions}
\label{sec:conc}

The maximum exhibited 
by $\langle{\cal P}\rangle$ against varying
$\mu$, shown in panel (a) of Fig.~\ref{fig:pol_cond_375}, can 
be explained with the help of Eqs.~(\ref{eq:gsplit})
and~(\ref{eq:pol_def2}), and Fig.~\ref{fig:pol_comp}. The
explanation given in Section~\ref{sec:spol} works because:
\par\noindent
(i)\ $\delta g_+$, $\delta g_-$ vanish continuously with increasing $\mu$,
for $\Delta \geq 0$;
 \par\noindent
(ii)\  $\delta^2g \equiv \delta g_+ -\delta g_->0 $ for $\Delta >0$ and 
suitably small $\mu$.

We do not know of a rigorous proof for either of statements (i) or (ii).
A plausibility argument for (i) is that since both $\delta g_+$
and $\delta g_-$ are positive for $\mu=0$, and given the spin-switching
character of $\mu$, the alternative for their vanishing with growing $\mu$
would be their turning negative. This would be at odds with the
notion of SO interactions as inducing channel equalisation. As for (ii),
we recall that (1) for  $\Delta  >0$ one should have $g_{++} > g_{--}$;
(2) the data in panel (b) of fig.~\ref{fig:pol_comp} show   that the 
{\em fractional} variations $\delta g_+/g_{++}$ and  
$\delta g_-/g_{--}$ are roughly equal
at any  given $\mu$, which appears reasonable. Points (1) and (2) 
help one understand (ii).

In summary, the maximum of $\langle{\cal P}\rangle$ against 
intensity of SO interactions is structurally linked to the behaviour
of spin-flipping and spin-conserving conductances, and some of their
combinations, namely $\delta g_+$, $\delta g_-$, and the second-order
difference $\delta^2g$. As this is a phenomenon taking place
in spin space, such feature is expected to be always present, irrespective
of the detailed form taken by the SO couplings, system size, and
field strength. 

The spin-dependent parameter
$\Delta$ added to the Hamiltonian in Sec.~\ref{sec:spol}
breaks time-reversal symmetry. So there is
no guarantee that results pertaining to that Section should exhibit
conservation of spin polarisation~\cite{nazarov}.
Nevertheless,
we have found for moderately strong site disorder $W$, and
for both weak and strong SO couplings, that the average spin
polarisation is kept constant along significant distances within the
system (though of course, 
variations associated with a very large
characteristic length cannot be ruled out). 
Moreover, in the special
case of low site disorder and weak SO interactions, we have evinced
spin-filter behaviour (see Fig.~\ref{fig:pol_mu02_w0_mvar}), with
spin polarisation increasing with distance from the injection edge.

The classical model  studied in~\cite{dqrbs17},
where SO effects are mimicked by so-called "spin-flipping" sites,
always exhibits polarisation {\em decay} against distance.
This would be in line with entropic interpretations of a spin-flipping
agent contributing towards equalising spin populations in an
initially spin-ordered beam.
In contrast with that, here we have not found  combinations of parameters 
which produce such behaviour.
So one sees that the symmetries imbedded in the quantum-mechanical
formulation used here are enough to counteract any such trend,
even though the initial beams are prepared in as close a manner as
possible in either case.

Investigation of the spin-spin correlation function~\cite{kka08},
see Fig.~\ref{fig:corf_mu02_w0_mvar}, yields
significant information on the characteristic decay length $\Lambda_s$,
for low and moderate site disorder,
provided the SO coupling is weak. For strong SO interactions
correlation decay takes place over short distances, comparable to
the lattice spacing.
It would then appear that the polarisation results for the classical process
studied in~\cite{dqrbs17}, and their always-decaying behaviour
against distance, find a closer qualitative correspondence here with the 
properties of the  correlation function, rather than those of its namesake.

The author thanks F Pinheiro, C Lewenkopf, T Ohtsuki, and J T Chalker
for helpful discussions, and the Brazilian agencies
CNPq  (Grant No. 303891/2013-0) and FAPERJ (Grant No. E-26/102.348/2013)
for financial support.


\end{document}